\documentclass{article}
\usepackage{authblk}
\usepackage{array}
\usepackage{listings}
\usepackage{graphicx}
\usepackage{bm}
\usepackage{color}
\usepackage{amsmath,amsfonts,amssymb}  
\usepackage[english]{babel}
\usepackage{booktabs,longtable}

\title{Electrical signature of individual magnetic skyrmions in multilayered systems}

\author[1]{Davide Maccariello}
\author[1]{William Legrand}
\author[1]{Nicolas Reyren}
\author[1]{Karin Garcia}
\author[1]{Karim Bouzehouane}
\author[1]{Sophie Collin}
\author[1]{Vincent Cros\thanks{Corresponding author: vincent.cros@cnrs-thales.fr}}
\author[1]{Albert Fert}

\affil[1]{Unit\'e Mixte de Physique, CNRS, Thales, Univ. Paris-Sud, Universit\'e Paris-Saclay, 91767, Palaiseau, France}

\date{}

\begin{document}

\maketitle

\begin{abstract}
Magnetic skyrmions are topologically protected whirling spin textures that can be stabilized in magnetic materials in which a chiral interaction is present. Their limited size together with their robustness against the external perturbations promote them as the ultimate magnetic storage bit in a novel generation of memory and logic devices. Despite many examples of the signature of magnetic skyrmions in the electrical signal, only low temperature measurements, mainly in magnetic materials with B20 crystal structure, have demonstrated the skyrmions contribution to the electrical transport properties. Using the combination of Magnetic Force Microscopy (MFM) and Hall resistivity measurements, we demonstrate the electrical detection of sub-100 nm skyrmions in multilayered thin film at room temperature (RT). We furthermore analyse the room temperature Hall signal of a single skyrmion which contribution is mainly dominated by anomalous Hall effect.

\end{abstract}

The electrical detection of skyrmions in lattice phase has been achieved only at low temperature, mainly in magnetic materials with B20 crystal structure \cite{Neubauer_2009, Lee_2009, Kanazawa_2011,schulz_2012,Li_2013,Porter_2014,du_2015}. Notably, contributions to the electrical transport properties has been directly linked to topological nature of skyrmions. More recently a change of the differential tunnel conductance due to the non-collinearity of single magnetic skyrmions has been reported for PdFe atomic bilayer on Ir(111) \cite{Hanneken2015,crum_2015}.
However, low temperatures preclude the use of magnetic skyrmions in real devices  \cite{Kiselev2011,Nagaosa2013,Fert2013}. The solution can be rendered by the engineering of magnetic multilayers in form of stacks composed of ultra thin magnetic layers in contact with heavy material thin films in which the stabilization of the magnetic skyrmions is achieved at room temperature  \cite{Jiang2015,Chen2015,Moreau-Luchaire2016,Boulle2016,Yu2016}. Although significant advances have been achieved in the stabilization and the current-induced motion of skyrmions at room temperature, only imaging techniques have been used \cite{Jiang2015,Woo2016,Jiang2016,Hrabec_2016,legrand_2017}. Importantly, no evidence of their detection through purely electrical means has been reported yet which is a crucial prerequisite for the potential applications. With this aim, we focus here on the transverse Hall resistivity, as a sensitive probe for the electrical characterization of magnetic skyrmions in patterned multilayered ultra thin films. 
The transverse resistivity $\rho_{xy}=V_y t/\,I_x$ is associated to the voltage $V_y$ measured across an electrical conductor (of thickness $t$), orthogonal to an applied current $I_{x}$. In magnetic materials with non-trivial magnetic texture, this transverse resistivity is usually decomposed in three components \cite{Nagaosa2010}: $\rho_{xy}=\rho_{xy}^{\rm OHE}+\rho_{xy}^{\rm AHE}+\rho_{xy}^{\rm THE}$, contributions arising from the ordinary, the anomalous and the topological Hall effect, respectively. The ordinary Hall effect (OHE) is described by $\rho_{xy}^{\rm OHE}= R_{\rm 0} H_\perp$, where $R_{\rm 0}$ is the ordinary Hall coefficient and $H_\perp$ the out-of-plane magnetic field component. The anomalous Hall effect (AHE), that shall contain intrinsic and extrinsic (negligible in our experiments) contributions arising from spin-orbit interactions, can be observed only in materials with broken time reversal symmetry \cite{Nagaosa2010}, as it is the case for ferromagnets (FM). The AHE resistivity $\rho_{xy}^{\rm AHE}$, for geometries considered here, is generally considered as being proportional to the average out-of-plane magnetization in the sample, $m_z$. 
The third term is the topological Hall effect (THE) that results from the Berry phase acquired by spin-polarised carriers as they go across a topologically non trivial texture such as skyrmions. In the adiabatic approximation, i.e. assuming a strong coupling of the charge with the local spin, the topological contribution is generally expressed by $\rho_{xy}^{\rm THE}=PR_{\rm 0}B_{eff}$, where $B_{eff}$ is the emergent ficticious field in the rest-frame of the electrons, and $P$ is the spin polarization  \cite{nagaosa_2013}.

As we recently demonstrated \cite{Moreau-Luchaire2016}, chiral magnetic skyrmions can be stabilized in our multilayers at room temperature by a large Dzyaloshinskii--Moriya interaction (DMI) arising at the interfaces of a thin Co layer inserted between two non-magnetic materials. For the present study, we have selected two different perpendicularly magnetized films based on the asymmetric trilayers Pt$|$Co$|$Al$_2$O$_3$ and Pt$|$Co$|$Ir, for which large values of DMI, ranging from 1 to $2$ mJ m$^{-2}$, \cite{Hrabec_2014,Yang_2015,Pizzini_2014,Belmeguenai_2015} allows skyrmions to be stabilized \cite{Moreau-Luchaire2016,schott_2016} resulting from of the balance between DMI, magnetic anisotropy and exchange coupling \cite{Fert2013,Bogdanov2001} as well as the dipolar energy. In the present work, we combine electrical transport measurements with local probe magnetic force microscopy (MFM) imaging, both realized at room temperature. A first objective is to control the generation of small magnetic skyrmions, using large current density pulses in e-beam lithographed structures (see Method for details). As reported by Romming et al. \cite{Romming_2013}, magnetic skyrmions can be generated by the injection of spin polarized currents through STM tip. Inhomogeneous in-plane electrical currents, created by geometrical constrictions, can also form bubble skyrmions due to the divergence of the spin torque acting on the magnetic domains walls, as shown by RT magneto optical Kerr miscroscopy \cite{Jiang2015, Olle_2016}.  
Recently, we have reported about the thermally assisted nucleation of isolated skyrmions at RT from the non-uniform magnetization states in micrometer sized tracks \cite{legrand_2017}. Here we show how the latter mechanism allows the generation of a finite number of randomly distributed magnetic skyrmions starting from the magnetically saturated state.

\begin{figure}[b]
{\includegraphics[width=11cm]{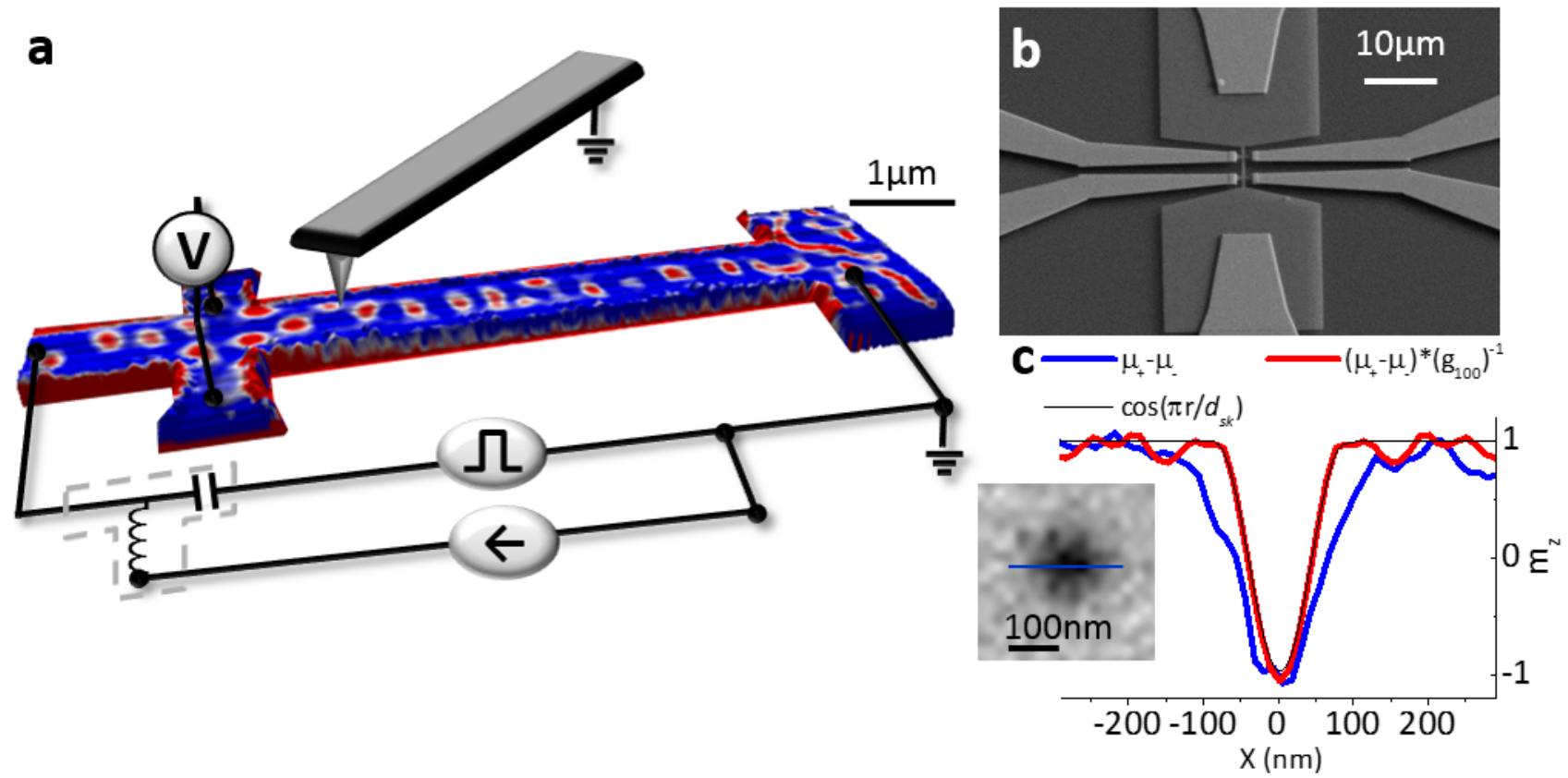}}
\caption{\textbf{Setup for skyrmion electrical detection and imaging}. \textbf{a}, Scheme of the experimental setup for electrical measurement integrated in a magnetic force microscope. Blue (red) color indicates attractive (repulsive) force on the MFM cantilever and hence reveal the magnetization in the sample. The sample shape is its actual topography measured by atomic force microscopy. \textbf{b}, Scanning electron microscopy image of a typical device. \textbf{c}, Scanning X-ray transmission microscopy (STXM) dichroic image of a skyrmion acquired at the Co $L_3$-edge in a sample of Ta(5)$|$Pt(10)$|$[Co(0.8)$|$Ir(1)$|$Pt(1)]$_{20}|$Pt(3) (number in parenthesis are thicknesses in nm). The profile of the dichroic signal in the STXM image (blue line, average over five scan lines through the skyrmion core) is used to determine the actual skyrmion out-of-plane magnetization profile $m_z$ (red line, see Methods) which is well approximated by the function $cos(\pi r / d_{sk})$ for $r<d_{sk}$ (black line).}
\label{fig1}
\end{figure}

In Fig.\ref{fig1}a, we show a schematic view of the experimental setup which allows us to directly relate the Hall resistivity response with the local magnetic configuration in the sample track: the topography rendered in perspective is colored with the MFM phase map, which is probing long range forces at 30\,nm from the surface. The magnetization state obtained after a finite number of current pulses as described hereafter, is characterized by the presence of a discrete number of skyrmions occupying the track of the patterned sample. The nature of the skyrmions and their size were previously investigated, in particular using scanning X-ray transmission microscopy \cite{Moreau-Luchaire2016} (STXM). The typical profile displayed in Fig.\ref{fig1}c allows us to determine (see Methods) the actual diameter $d_{sk}$ of the skyrmions that we will consider in the rest of the article and is defined as the diameter of the circle on which $m_z$ = 0.

\begin{figure}[!h]
{\includegraphics[width=9 cm]{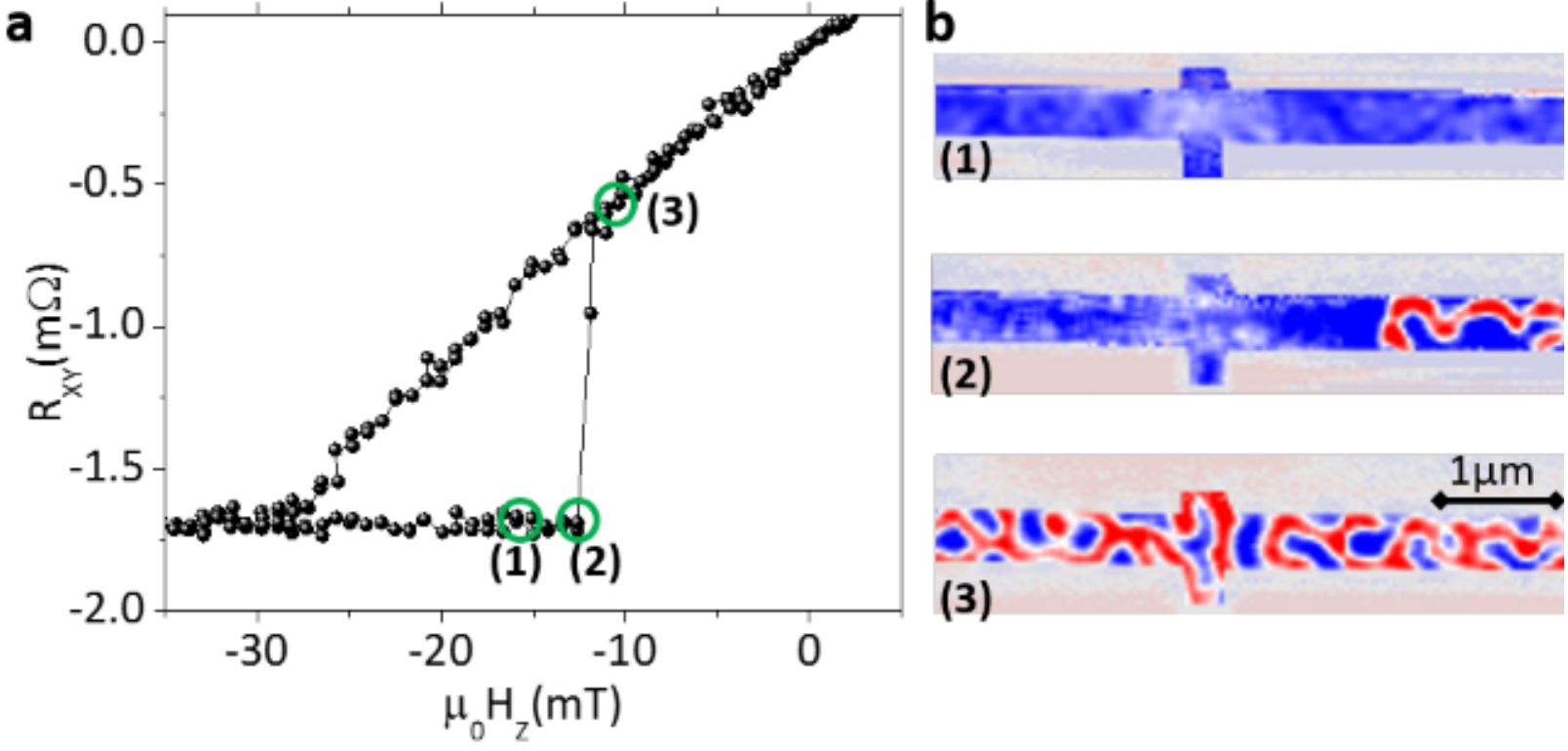}}
\caption{\textbf{Hall resistance vs magnetic configuration in the Hall cross}. \textbf{a}, Hall resistance and nucleation of magnetic domain in a selected region of out-of-plane field hysteresis loop. The MFM phase maps (right panel) shows that the transverse resistance $R_{xy}$ (left graph) is constant (point \textit{B}) until magnetic domains enter the Hall cross. In \textit{(2)} the MFM image evidences the nucleation of worm domains in the track far from the Hall cross. $R_{xy}$ exhibits variations only when the the magnetization evolution interests the Hall cross area ($3$).} 
\label{fig2}
\end{figure}

The accuracy of the electrical probe of the local magnetic configuration in the Hall cross area is shown in Fig.\ref{fig2}: The transverse resistance \textit{R}$_{xy}$ is plotted as a function of $H_\perp$ in a track-shaped Pt(10)$|$[Pt(1)$|$Co(0.6)$|$Al$_{2}$O$_3$(1)]$_{5}$ sample (in parentheses are given the thicknesses in nanometers). Starting from large negative field, the uniform magnetic state is assessed by both the MFM image and the saturation of the Hall resistance. The state is preserved when the absolute value of the field is decreased (point \textit{A}, still $H_\perp<0$). If the magnetic field is set at the value of the point \textit{B}, the Hall resistance remains constant while the MFM phase map shows the onset of worm-like magnetic domains in the track, still about 1\,$\mu$m away from the center of Hall cross. The variation of $R_{xy}$ can be observed only when a magnetic domain approaches or is nucleated in the Hall cross area (point \textit{C}): After a sudden change of $R_{xy}$ associated with the appearance of a first reversed magnetic domain in the Hall cross area, the transverse resistance then smoothly evolves with the size of magnetic domains growing in the Hall cross.

\begin{figure}[!h]
{\includegraphics[width=0.9\textwidth]{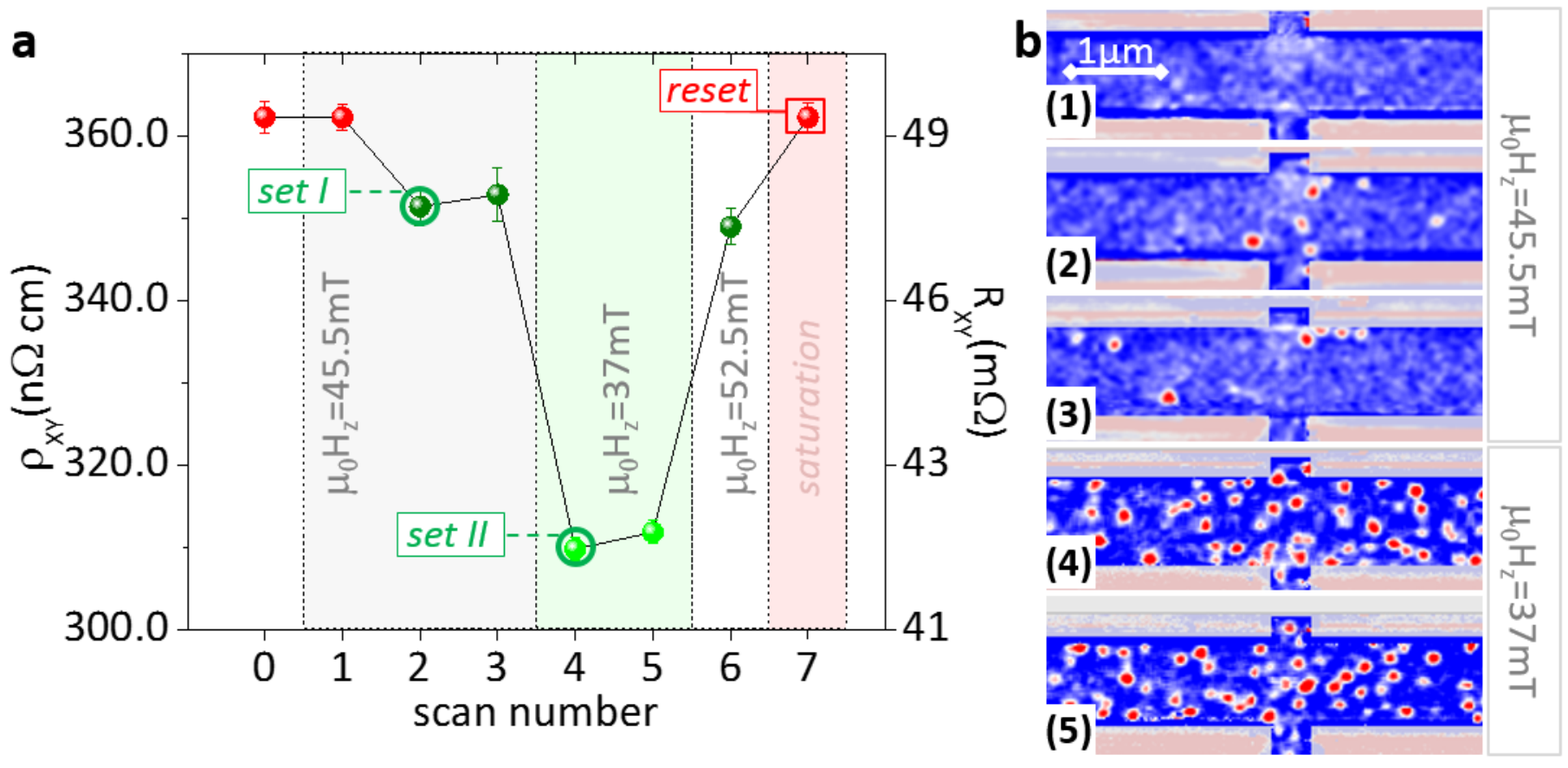}}
\caption{\textbf{Room temperature detection of current nucleated skyrmions}.\textbf{a}, Variation of the Hall resistivity (left-hand side scale) and Hall resistance (right-hand side scale) as a consequence of the nucleation of skyrmions by current pulses at different magnetic fields (colored areas). \textbf{b}, MFM phase map related to Hall resistivity shown in a. The numbers labeling the images correspond to the $x$-axis of a. These results clearly show that the number of skyrmions hosted in the track after pulses are sent depends on the applied magnetic field as demonstrated by the three different levels indicated by \textit{set I}, \textit{set II} and \textit{reset}. Moreover, it is shown that the initial state can be recovered by increasing $H_\perp$ to the saturation, indicated with \textit{reset}. } 
\label{fig3}
\end{figure}

Hence, we use such measurements of Hall voltage to detect the local small changes in the magnetic configuration and to demonstrate that isolated skyrmions can be electrically detected in tracks with dimensions approaching the skyrmions size. In order to reproducibly nucleate skyrmions from a uniform magnetic state, our strategy consists in first saturating the sample magnetization by applying a large out-of-plane magnetic field, then decreasing its absolute value to a selected value before applying current pulses to create skyrmions. The variations of $R_{xy}$ in a patterned sample with composition Ta(5)$|$Pt(10)$|$[Co(0.8)$|$Ir(1)$|$Pt(1)]$_{20}|$Pt(x) after application of a discrete number of current pulses are displayed in Fig.\ref{fig3}. 
After saturating of the magnetization, obtained by applying an out-of plane field of $140$\,mT, we start with a perpendicular field of $45.5$\,mT.  Both MFM and transport measurement confirm that the sample preserve the uniform magnetic state as shown by the scan number \textit{(1)}. In this situation and the similar uniform state \textit{(0)}, the Hall resistivity is found to be around 360 n$\Omega$\,cm.
 The former field value has been chosen in the range in which, previous field-loop MFM characterization shows that after the  magnetization reversal, the magnetic configuration presents only skyrmions and no elongated domains. When a set of pulses ($n=20$), having a duration of $100$\,ns and current density of $J=2 \times 10^{11}$\,A\,m$^{-2}$, is applied to the track, the nucleation of a small number of skyrmions (scan number \textit{(2)}) is observed. The transverse resistivity $\rho_{xy}$ likewise shows a small drop related with the appearance of the skyrmions in the cross-bar (named \textit{set I}).
Further current pulses change the skyrmions distribution in the track but the skyrmions density keeps constant. Indeed, after several sequences of pulses we observe, in the MFM images, the appearance or disappearance of skyrmions at different positions of the track. Accordingly, we record only a small variation of $\rho_{xy}$ that is related with the small variation of number of skyrmions in the area of the cross-bar. Even after 500 pulses (scan number \textit{(3)}), we observe that the magnetic configuration does not appreciably change, i.e. without any nucleation of worm domains and with a similar total number of skyrmions in the full track.

To further modify the skyrmion density, the applied out-of-plane magnetic field $H_\perp$ must be changed: The current induced nucleation of a larger skyrmion density is obtained after decreasing the field (in absolute value),  i.e. to 37\,mT (point \textit{(4)} in Fig.\ref{fig3}). The appearance of this dense ensemble of skyrmions occupying the whole track, is characterized by a large drop of $\rho_{xy}$ (\textit{set II}, displayed in Fig.\ref{fig3}a). As described earlier for a larger field, successive current pulses only slightly change the observed skyrmions density in the MFM images and the $\rho_{xy}$ keeps a nearly constant value (point \textit{(5)}). The track can hence be thought as a skyrmion reservoir being in a saturated state with a number of hosted skyrmions depending on the magnetic field. The previous state can be reestablished by increasing the out-of-plane field (see point \textit{(6)}), and finally the initial state can be achieved again ("\textit{reset}") if the magnetic field is increased up to the magnetic saturation where the MFM image (not shown here) and the transverse resistivity around 360 n$\Omega$\,cm (point \textit{(7)}) corresponds to the uniform magnetization state. Moreover, we also found that at larger field, {\it i.e.} $\mu_0H_\perp=57$\,mT, when skyrmions are still present, the reset can be also obtained by applying a few current pulses having comparable current densities as the ones used before. This observation indeed proves that the small skyrmions that are still metastable under a large field, can be easily erased by the temperature rise originating from Joule heating: The energy barrier avoiding skyrmions to collapse into the uniform state is now overcome by the thermal energy.

\begin{figure}[!h]
{\includegraphics[width=1.1\textwidth]{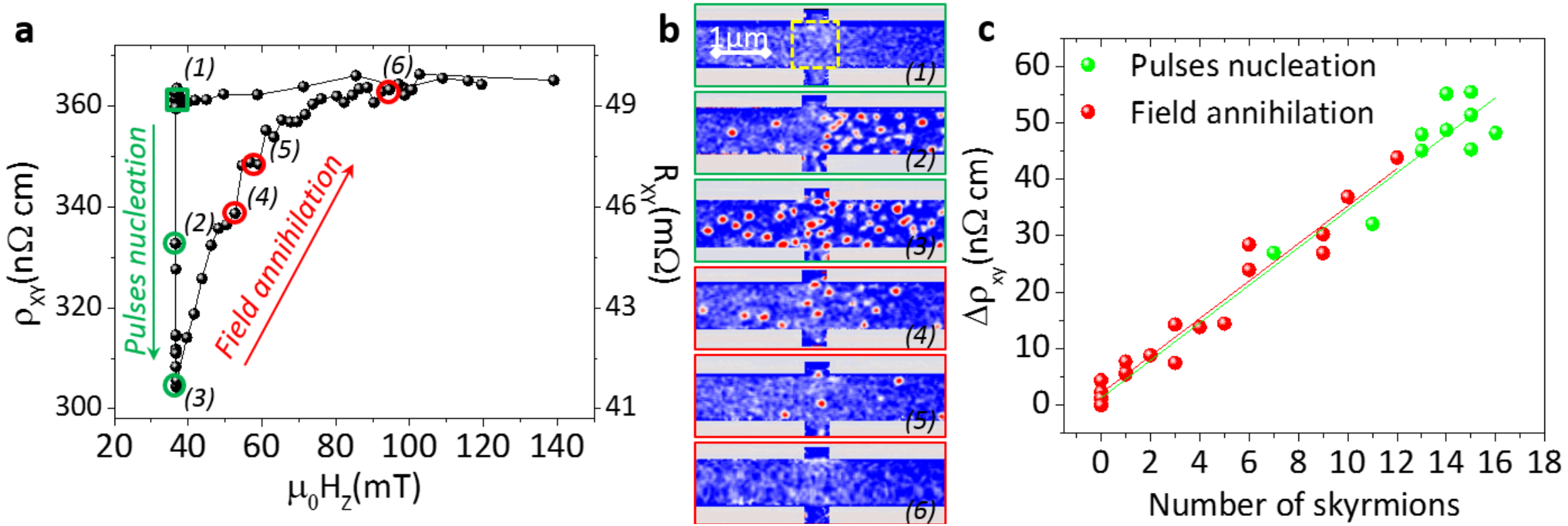}}
\caption{\textbf{Room temperature transverse resistivity due to single skyrmions}. \textbf{a}, The current pulse assisted nucleation is realised at $37$\,mT, after magnetic saturation at positive field. The skyrmions nucleation leads to a variation of $\rho_{xy}$. On the contrary, by slowly increasing the applied field we assist to the skyrmion annihilation. \textbf{b}, The variation of the transverse resistance is highly sensitive to the number of skyrmions in the track as shown by selected MFM images. \textbf{c}, Resistivity change $\Delta\rho_{xy} = (R_{xy}({\rm saturated})-R_{xy})\,t_{\rm tot}$ plotted as function of the number of skyrmions in the Hall cross area defined by the dotted rectangle in b. The nucleation (green dots) and the annihilation process (red dots) exhibit the same slope ${\rm d}\Delta\rho_{xy}/{\rm d}N$. The linear fit to the experimental data (red lines) gives $\Delta\rho_{xy}^{sk} = 3.5\pm 0.5$\,n$\Omega$\,cm for a single skyrmion.}
\label{fig4}

\end{figure}

Having determined a method to nucleate and detect magnetic skyrmions in a track, we now estimate the RT signal $\Delta\rho_{xy}$ corresponding to the single skyrmion. The system is first magnetically prepared in the uniform state, then skyrmions are introduced using the strategy described above, choosing a starting field $\mu_0H_\perp= 37$\,mT and sending single current pulses with the same properties than above: $J=2 \times 10^{11}$\,A\,m$^{-2}$ and $100$\,ns duration. We then monitor the evolution of the magnetic configuration by MFM and measure $\rho_{xy}$ concomitantly after each pulse: the evolution of $\rho_{xy}$ during this current-induced nucleation processes at a fixed out-of-plane magnetic field is displayed in Fig.\ref{fig4}a with label \textit{pulse nucleation}. Thereafter, we further monitor through MFM and electrical transport measurements the annihilation process achieved by slowly increasing the field (indicated by \textit{field annihilation} in Fig.\ref{fig4}a). The progressive suppression of skyrmions (at different critical fields) might be enhanced due to inhomogeneities such as the variation of local DMI, magnetic anisotropy or exchange stiffness observed, in similar asymmetric multilayers \cite{legrand_2017,kim_2017,Bacani_2016}. In this scenario,  magnetic inhomogeneities lead to a broad distribution of the skyrmions annihilation field. After magnetic saturation, the shape of the initial magnetic loop of $\rho_{xy}$ can be recovered and the experience can be reproduced: this reversibility demonstrates that the current pulses are not affecting the material magnetic properties but only the the magnetic texture.

In Fig.\ref{fig4}b we display representative MFM images for both pulse nucleation (green frame) and field annihilation (red frame) corresponding to the points highlighted in the transport measurement. Note that, for the selected fields, the current nucleation process only allows the generation of magnetic skyrmions and no elongated domain can be observed as it is the case for lower applied magnetic fields (not shown). 
Selecting a portion of the track restricted to the Hall cross region and lateral size of the length scale of the track width (see Fig.\ref{fig4}b), it is possible to count the number $N$ of skyrmions and plot the variation of $\Delta\rho_{xy}$ as function of it (Fig.\ref{fig4}c). Both nucleation and annihilation processes exhibit the same trend in the transverse resistivity: variations of $\rho_{xy}$ are only due to the varying number of skyrmion in the Hall cross area. The linear regression of our experimental data, for both processes, indicates that each skyrmion contributes equally to $\rho_{xy}$ with an amount $\Delta\rho_{xy}^{sk} = 3.5\pm0.5$\,n$\Omega$\,cm. One should also note that the skyrmion diameter (for our films) is almost not varying anymore at high field before annihilation explaining the identical slope in both nucleation and annihilation experiments. With $N=15$ skyrmions, the total resistivity change $\Delta\rho_{xy}$ increases up to a value of about $58$ \,n$\Omega$\,cm.

In order to understand the origin of the variation of the Hall signal with the change of the number of skyrmions, we compare the experimental value with those expected for both AHE and THE contributions. The skyrmion area $a_{sk} = \pi d_{sk}^2/4$ is a key parameter that we have better estimated using STXM imaging, which is sensitive to the out-of-plane magnetization, and not from the MFM images, in which the skyrmion size might be affected by the interaction with the magnetic tip. As already shown in Fig.\ref{fig1}c, in the range of magnetic field of these experiments, the skyrmion diameter is $d_{sk}=85\pm 10$\,nm, similarly to what has been found in similar magnetic thin film based on the Pt$|$Co$|$Ir trilayer \cite{Moreau-Luchaire2016}.

The magnetization profile in Fig.1c can be described by $m_z(r)$ = $m$ $cos(\pi r / d_{sk})$, where \textit{m} is the value of $m_z$ outside the skyrmion and $r$ the distance from the centre. Consequently, the mean value of $m_z$ inside the skyrmion is approximately $4 m/\pi^2$. If $n$ is the skyrmion density  per unit area of the track in the vicinity of the Hall contacts, the mean magnetization per unit area and the Hall resistivity are reduced by the factor (1 - $16 n a_{sk}/\pi^2$). The contribution of a single skyrmion can be written as 
$$\Delta \rho^{\rm AHE} \approx \rho^{\rm AHE} 16 a_{sk}/ \pi^2 a_{cross} $$
where $a_{cross}$ is the area of the track in which the distribution of the transverse electric field is affected by the skyrmion (the affected length of the track length  is of the order of its width). 
Taking $\rho^{\rm AHE}_{xy}= 360  $\,n$\Omega$\,cm from Fig.3, $a_{cross}$= 1.0 $\pm\, 0.1$ $\mu m^2$ and $a_{sk} = (5.7 \pm\, 0.7) \times 10^{-3}$  $\mu m^2$, we obtain $\Delta\rho^{\rm AHE}_{xy}\approx3.3\pm\,0.5 $\,n$\Omega$\,cm which is in excellent agreement with the measured one. The estimation of the discrete THE contribution of a single skyrmion is still somehow controversial because so far this contribution has only been observed in systems with dense skyrmion latticesin B20 non centrosymmetric systems \cite{Neubauer_2009, Lee_2009, Kanazawa_2011,Li_2013}. 
These materials show also low electron densities and a strong exchange interaction justifying the adiabatic approximation. The unit topological Hall resistivity related to the integer winding number of a skyrmion,  for a cross bar of width $a_{cross}$, can be assumed to be $\rho_{xy}^{\rm THE}=-P R_{\rm 0} \Phi_0/a_{cross}  $ \cite{kanazawa_2015}
, where $\Phi_0$ is the flux quantum. A typical value of the spin polarization for Co is $P=0.4$, whilst the experimental Hall coefficient that we have measured for our sample is $R_{\rm 0}=2\times 10^{-11}$ $\Omega$ m T$^{-1}$. The estimated THE contribution for single skyrmion, if we also ignore any possible reduction by a too short mean free path, is hence $\rho_{xy}^{\rm THE}=(1.7$ $\pm$ $0.1) \times 10^{-3}$\,n$\Omega$\,cm, that is three order of magnitude lower than the observed value. Consequently, we mainly attribute the observed $\Delta\rho_{xy}^{sk}$ to the AHE related to our skyrmions and estimate that THE plays a negligible role in our experiment with diameters around 100 nm. 

Although the variation of the Hall resistance that we report in this work is of the same order of magnitude than those observed for the THE at low temperature in MnSi \cite{Neubauer_2009}, we ascribe the electrical signal to a different origin. Unlike B20 materials, in our metal multilayers, with skyrmions in the range of 100 nm, AHE is the dominant skyrmion contribution to the transverse electrical signal. The main difference could lie in the fact that in B20 materials it is possible to stabilize dense skyrmion lattices with very short magnetic modulation period of about $3$\,nm for MnGe \cite{Kanazawa_2011}. The emergent magnetic field $B_{eff}$ in these materials can reach high values, \textit{e.g.}  11 T for MnSi \cite{Neubauer_2009}, and the THE is dominant \cite{Nagaosa2013}.
On the contrary, in metallic multilayers the low number of skyrmions (with size of about 85\,nm) combined with the small ordinary Hall coefficient of metals makes the topological contribution negligible compared to AHE. 
In addition, the reduced electron mean free path with respect to the skyrmions size, as well as the large charge carrier density of the metallic systems, shall result in a significant reduction of the topological contribution term of skyrmions to the Hall resistance \cite{Hamamoto_2016}. 
Recently, a transition from charge to spin topological Hall resistance has been theoretically predicted if the exchange interaction varies \cite{denisov_2017}. The dependence of the THE on the exchange splitting and skyrmion size denotes that the estimation of the topological contribution might be non trivial \cite{denisov_2017}.

Keeping in mind the targeted requirements for devices, the skyrmion diameter should probably be reduced in size about three folds, and the track adjusted to the skyrmion size. In this case, the ratio of the skyrmion area over the probe area being maximized, the AHE signal will be considerably increased: Assuming 30\,nm diameter skyrmion and similar electrical properties of the magnetic multilayer, in a 60\,nm wide track with Hall probe of the same width, $\Delta\rho_{xy}^{\rm AHE}$ would reach $4  \rho_{xy}^{\rm AHE}/\pi^2 \approx 197$\,n$\Omega$\,cm. The THE on the contrary could reach about $5$\,n$\Omega$\,cm, still one order of magnitude smaller than the AHE contribution.

In summary, we demonstrate in lithographed magnetic multilayers that, starting from a uniform magnetic state, we can electrically nucleate a discrete number of skyrmions and electrically detect their presence. Tuning the applied magnetic field and the current pulse allows the skyrmion density in tracks to be controlled. The transverse resistivity, mainly dominated by the anomalous Hall contribution, scales linearly with the number of skyrmions in the track. This allows to detect and count the skyrmions by electrical means. The room temperature observation of skyrmions and their electrical detection strongly advances the technological perspective for room temperature skyrmion-based devices and it can represent a new avenue for further fundamental studies on their very rich physics.

\section*{Methods}
The multilayers samples were grown at room temperature by dc sputtering on thermally oxidized Si substrates. Hall cross-bars geometries were defined by electron beam lithography and Ar ion etching. The electrical transport characterization was performed by using a nanovoltmeter to measure the transverse ''Hall'' voltage with a dc current provided by a current source. The current density used for the dc transport characterization is $J \approx 2\times 10^{8}$ A m$^{-2}$, way below the nucleation current density. The MFM images confirm that the magnetic configuration is not altered by the Hall voltage measurements.
The current-induced nucleation of skyrmions is obtained by short current pulses ($t_p = 30-200$\,ns) applied by an arbitrary signal generator through a bias-tee in order to isolate the dc and ac circuits. The applied pulses were separated by more than 1\,ms in order to allow the dissipation of the Joule heating. The local magnetic configuration in the sample was probed using an MFM setup in "lift mode'' at a height of 30\,nm over the surface. The MFM imaging is performed before and after the measurements of the Hall resistance in order to check that the domains conformation is stable. To accurately estimate the skyrmion diameter, we performed scanning transmission X-ray microscopy (STXM) scans on samples deposited on 200-nm-thick Si$_3$N$_4$ membranes. The profile of the skyrmion magnetization is evaluated by deconvolution of the XRMS profile with a Gaussian-shaped beam $g_{100}$ having a full width at half maximum (FWHM) of 100\,nm (effective value estimated by different means): The STXM dichroic profile $\mu_+-\mu_-$ is assumed to be the convolution of the magnetization $m_z$ of the real skyrmion with the beam profile. By the theorem of the convolution the profile of the $m_z$ of the skyrmion is found by $m_z$ = FT$^{-1}$[FT($\mu_+-\mu_-$)/FT($g_{100}$)]. The skyrmion size ($d_{sk}=85\pm 10$\,nm) is then defined by the contour $m_z$ = 0. 

\section*{Acknowledgement}
The authors acknowledge C. Moreau-Luchaire for participating to sample preparation, A. Vecchiola for the technical support in the MFM measurements and C. Moutafis, S. Finizio, P. Warnicke and J. Raabe for their technical support at the (PolLux) beamline at the SLS, Paul Scherrer Instit\"{u}t, Villigen, Switzerland.
The authors acknowledge financial support from European Union grant MAGicSky No. FET-Open-665095.

\section*{Author contributions}
N.R., V.C. and A.F. conceived the project. S.C. grew the multilayer films. K. G. patterned the samples. D.M. acquired the data of MFM, transport measurements and STXM, treated and analysed the data with the help of N.R., W.L., K.B. and V.C.. D.M., N.R., V.C. and A.F. prepared the manuscript. All authors discussed and commented the manuscript.

\bibliography{References}

\end{document}